\documentclass[aps,prl,twocolumn,superscriptaddress]{revtex4-1}

\usepackage{graphicx}
\usepackage{amsmath}
\usepackage{color}

\usepackage[bookmarks=true,colorlinks=true,urlcolor=blue,linkcolor=blue,citecolor=blue,breaklinks]{hyperref}

\usepackage{amssymb}
\usepackage{booktabs}

\begin{document}

\sloppy

\title{Ultrafast Hot Phonon Dynamics in MgB$_2$ Driven by Anisotropic Electron-Phonon Coupling}

\newcommand*{\DIPC}[0]{{
Donostia International Physics Center (DIPC),
Paseo Manuel de Lardizabal 4, 20018 Donostia-San Sebasti\'an, Spain}}

\newcommand*{\IFS}[0]{{
Center of Excellence for Advanced Materials and Sensing Devices, Institute of Physics, Bijeni\v{c}ka 46,
10000 Zagreb, Croatia}}

\newcommand*{\ISMCNR}[0]{{
Istituto di Struttura della Materia, CNR,
Division of Ultrafast Processes in Materials (FLASHit),
34149 Trieste, Italy}}

\newcommand*{\HU}[0]{{
Institut f\"ur Physik and IRIS Adlershof, Humboldt-Universit\"at zu Berlin, Berlin, Germany}}

\author{D. Novko}
\email{dino.novko@gmail.com}
\affiliation{\IFS}
\affiliation{\DIPC}

\author{F. Caruso}
\affiliation{\HU}

\author{C. Draxl}
\affiliation{\HU}

\author{E. Cappelluti}
\email{emmanuele.cappelluti@ism.cnr.it}
\affiliation{\ISMCNR}


\begin{abstract}
The zone-center $E_{2g}$ modes play a crucial role
in MgB$_2$, controlling the scattering mechanisms
in the normal state as well the superconducting pairing.
Here, we demonstrate 
via first-principles quantum-field theory calculations
that, due to the
anisotropic electron-phonon interaction, a {\em hot-phonon} 
regime where the  $E_{2g}$ phonons can achieve significantly larger effective populations
than other modes,  is triggered in MgB$_2$ by the interaction with an 
ultra-short laser pulse.
Spectral signatures of this scenario
in ultrafast pump-probe Raman spectroscopy are discussed in detail,
revealing also a fundamental role of nonadiabatic processes
in the optical features of the $E_{2g}$ mode. 
\end{abstract}

\maketitle




Although MgB$_2$ is often regarded as a conventional
high-$T_c$ superconductor, described by the 
Eliashberg theory for phonon-mediated superconductivity, 
it displays many peculiar characteristics
that make it a unique case.
Most remarkable is the anisotropy of the electronic and superconducting
properties, where electronic states belonging to the
$\sigma$ bands are strongly coupled to phonons, and display thus
large superconducting gaps $\Delta_\sigma$, whereas
electronic states associated with the $\pi$ bands are
only weakly coupled to the lattice,
and hence exhibit small superconducting gaps $\Delta_\pi$ 
\cite{bib:liu01,bib:choi02b,bib:choi02a,kong,golubov,giubileo,tsuda,chen2001,gonnelli,mou}.
Such electronic anisotropy
is also accompanied by a striking anisotropy in the {\em phonon} states.
The electron-phonon (e-ph) coupling is indeed 
strongly concentrated in 
few in-plane $E_{2g}$ phonons modes
along 
the $\mathrm{\overline{\Gamma}-\overline{A}}$ path of the Brillouin zone \cite{kong,an,yildirim},
whereas the remaining e-ph coupling is spread over all 
other lattice modes in the Brillouin zone.

Due to its pivotal role in ruling e-ph based many-body effects
and in the superconducting pairing,
the properties of the 
long-wavelength $E_{2g}$ mode have been extensively investigated,
both theoretically and experimentally
\cite{yildirim,bohnen,hlinka,goncharov,postorino1,renker,bib:quilty02,bib:martinho03,bib:shi04,dicastro06,bib:shukla03,boeri02,profeta,bib:lazzeri03,bib:calandra05,boeri05,bib:cappelluti06,simonelli,novko18}.
On the experimental side, Raman spectroscopy 
has proven particularly suitable for 
providing fundamental information on the lattice dynamics
and on the many-body e-ph processes.
Particularly debated is the origin of the
large phonon linewidth
$\Gamma_{E_{2g}} \approx 25$\,meV,
and of the temperature dependence of both the phonon frequency
and linewidth \cite{yildirim,bohnen,hlinka,goncharov,postorino1,renker,bib:quilty02,bib:martinho03,bib:shi04,dicastro06,bib:shukla03,boeri02,profeta,bib:lazzeri03,bib:calandra05,boeri05,bib:cappelluti06,simonelli,novko18}.
The complexity of identifying the quantum-mechanical origin of these phenomena arises 
from the concomitance of the e-ph interaction, non-adiabaticity, and 
lattice anharmonicities, in turn responsible for phonon-phonon scattering 
and thermal expansion.
A possible path for tuning selectively only one of these processes
is thus highly desirable, in 
order to disentangle the different mechanisms in action.

Ultrafast time-resolved optical characterizations of MgB$_2$
with a pump-probe setup were presented in
Refs.\,\cite{xu03,demsar03,baldini}, where two different relaxation times
were identified in the normal states.
In particular, 
the observed anomalous blueshift at a short time scale
of the in-plane plasmon
was qualitatively explained in Ref.\,\cite{baldini}
by assuming that the $E_{2g}$ mode
behaves as a {\em hot} phonon, i.e., a lattice mode
with larger population compared
with the thermal distribution of the
other lattice degrees of freedom (DOFs), in analogy
with what was recently observed in graphite and
graphene
\cite{bib:yan09,bib:berciaud10,bib:chae10,butscher,wang,scheuch,huang,bib:wu12,bib:ferrante18}.
A similar scenario was suggested in Ref.\,\cite{demsar03}.
However, the actual observation of hot-phonon physics in MgB$_2$ was
quite indirect, and further compelling evidence is needed.

In this Letter we present
a detailed theoretical investigation of the time-resolved Raman spectroscopy
of the $E_{2g}$ mode in a pump-probe setup.
Using {\em ab-initio} and quantum-field-theory techniques,
we predict that non-equilibrium processes in
MgB$_2$ are dominated by strong hot-phonon physics.
Several detailed experimental characterizations
are suggested which can provide a direct and decisive evidence
of the hot-phonon dynamics. 
It is worth stressing that,
unlike graphene where the hot-phonon 
physics stems from the reduced phase space
available for e-ph scattering (due to the vanishing
Fermi area at the Dirac points) \cite{bib:yan09,bib:berciaud10,bib:chae10,butscher,wang,scheuch,huang,bib:wu12,bib:ferrante18},
the hot-phonon properties in MgB$_2$
are ruled by the strong anisotropy of the e-ph coupling,
with the most of the coupling strength being concentrated
in few phonon modes at the Brillouin zone center.
Such new theoretical paradigm for inducing hot-phonon
physics is not limited to MgB$_2$ but it is
quite general, and it can be applied to different materials
in order to elucidate the time-resolved infrared spectroscopy of 
the zone-center phonon modes in general. 
Our work paves the way for a direct experimental check of
hot-phonons in MgB$_2$ and in other similar materials characterized
by a strongly anisotropic e-ph coupling.

Density-functional theory calculations were performed by using the
{\sc quantum espresso} package\,\cite{bib:qe}.
Norm-conserving pseudopotentials were employed with the 
Perdew-Burke-Ernzerhof
exchange-correlation functional \cite{bib:pbe}.
A $24 \times 24\times 24$ Monkhorst-Pack grid in
momentum space and a plane-wave cutoff energy of 60\,Ry were
used for ground-state calculations.
The phonon dispersion was calculated on a
$12\times12\times12$ grid using 
density-functional perturbation theory (DFPT),\cite{bib:baroni01},
and the e-ph coupling was computed
by using an in-house modified version of the {\sc epw} code \,\cite{bib:epw}.
Electron and phonon energies, and
e-ph coupling matrix elements were interpolated using maximally-localized Wannier functions\,\cite{bib:wannier}. 
The phonon self-energy for the ${\bf q}=0$ $E_{2g}$ mode was computed on a $300\times300\times300$
electron momentum grid, while the Eliashberg function was obtained on a
$40\times40\times40$ grid of electron and phonon momenta.

%
\begin{figure}[!t]
\includegraphics[width=0.48\textwidth]{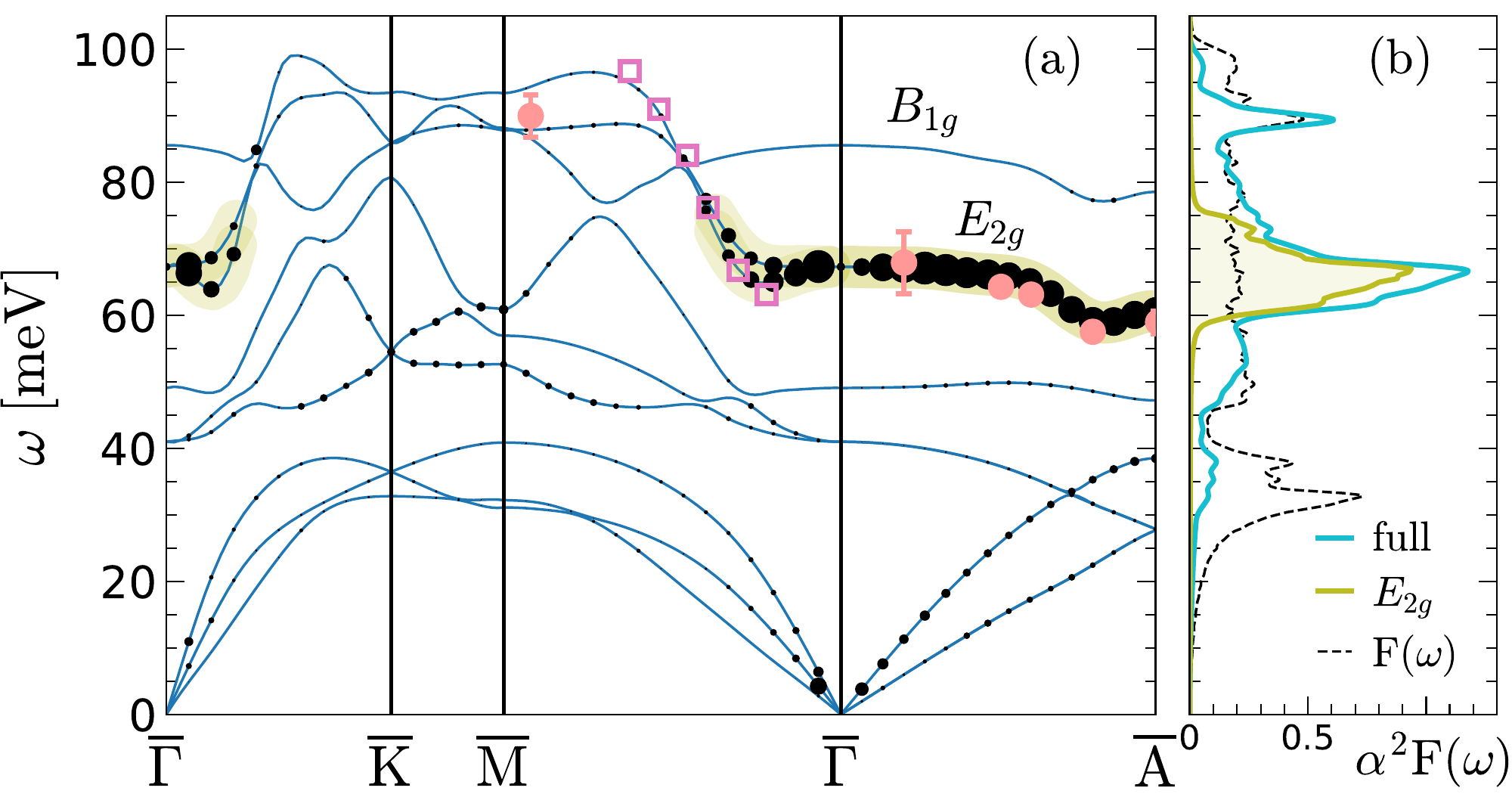}
\caption{\label{fig:fig1}(a) Plot of the
phonon dispersions (solid lines)
and e-ph coupling strengths $\lambda_{\mathbf{q}\nu}$,
represented by the size of the black circles.
Also shown are the experimental phonon
energies of the $E_{2g}$ mode close to the $\mathrm{\overline{M}}$ point and
along the $\mathrm{\overline{\Gamma}}-\mathrm{\overline{A}}$ path (red circles)\,\cite{bib:shukla03}, as well as along the $\mathrm{\overline{M}}-\mathrm{\overline{\Gamma}}$ cuts (purple empty squares)\,\cite{bib:baron04}.
(b) Corresponding
phonon density of states $F(\omega)$ (dashed line)
and the total Eliashberg function $\alpha^2F(\omega)$ (blue solid line).
Green color shows the contribution to the Eliashberg function
associated with the hot $E_{2g}$ modes around and along the $\mathrm{\overline{\Gamma}}-\mathrm{\overline{A}}$ path,
$\alpha^2F_{E_{2g}}(\omega)$.
}
\end{figure}
%

The phonon dispersion and the e-ph coupling strengths $\lambda_{\mathbf{q}\nu}$ are depicted in Fig.\,\ref{fig:fig1}(a),
and the corresponding phonon density of states
and Eliashberg function $\alpha^2F(\omega)$
in Fig.\,\ref{fig:fig1}(b).
Our computed phonon dispersions are in good
agreement with previous 
results\,\cite{bib:liu01,bib:choi02b,bib:choi02a,kong,golubov,bib:shukla03,bib:baron04,bib:eiguren08},
while the total e-ph coupling strength $\lambda=0.6$ 
is smaller than the earlier {\em ab-initio} values ($\lambda\gtrsim 0.7$)\,\cite{bib:liu01,kong,bib:eiguren08,bib:calandra10,bib:margine13}, but in 
rather good agreement with experimental estimates\,\cite{bib:bouquet01,bib:wang01}.
Consistently with earlier works \cite{kong,an,yildirim}, large values
of the e-ph coupling  
are mainly concentrated in the $E_{2g}$ branch
in the Brillouin zone center along the $\mathrm{\overline{\Gamma}-\overline{A}}$ line.
This is reflected in a dominant peak in
the Eliashberg function at the corresponding $E_{2g}$ energies
$\omega \approx 60-70$ meV.
As shown below,
such remarkable anisotropy is responsible for the
hot-phonon scenario, where the zone-center $E_{2g}$ phonon modes
can acquire,
under suitable conditions (i.e., by using pump-probe techniques), a
population much larger than other underlying lattice DOFs.

In order to capture the anisotropy of the e-ph interaction, 
we model the total Eliashberg function
as sum of two terms,
$\alpha^2F(\omega)=\alpha^2F_{E_{2g}}(\omega)
+\alpha^2F_{\rm ph}(\omega)$,
where $\alpha^2F_{E_{2g}}(\omega)$
contains the contribution of the hot $E_{2g}$ modes along and
around the $\mathrm{\overline{\Gamma}-\overline{A}}$ path
in the relevant energy range $\omega \in [60:75]$ meV
 (green shaded areas in Fig.\,\ref{fig:fig1}),
while $\alpha^2F_{\rm ph}(\omega)$ accounts
for the weakly coupled cold modes in the remnant
parts of the Brillouin zone.
The resulting e-ph coupling strengths for the hot and cold modes are
$\lambda_{E_{2g}}=0.26$ and $\lambda_{\rm ph}=0.34$, respectively.

With the fundamental input of the anisotropic e-ph coupling,
we investigate the rates of the energy transfer between the
electron and lattice DOFs
in a typical time-resolved pump-probe experiment.
As we detail below, energy transfer processes
and the hot-phonon physics are driven by the strong
anisotropy of the thermodynamical properties of hot and cold modes
i.e. by the remarkable difference in specific heats.
This physics do not rely thus on the assumption of
effective temperatures for the electronic and lattice DOFs.
On the other hand, the use of standard three-temperature model
appears as a reliable and convenient way to describe these processes
in terms of few intuitive
quantities\,\cite{bib:allen87,bib:perfetti07,lui2010,bib:dalconte12,bib:johannsen13}.
The validation of this modelling, compared with the results
of a numerical computation using {\em non-thermal} distributions,
is presented in Ref.\,\cite{bib:si} (for detailed comparison between thermal and non-thermal models see Section\,S2 and Figs.\,S2 and S3).
Characteristic parameters
of our description will be thus the effective electronic temperature $T_{\rm e}$,
the effective temperature $T_{E_{2g}}$ of the
hot $E_{2g}$ phonon strongly coupled to
the electronic $\sigma$ bands,
and the lattice temperature $T_{\rm ph}$ that
describes the effective temperature
of the remaining cold phonon modes:
\begin{eqnarray}
C_{\rm e}
\frac{\partial T_{\rm e}}{\partial t}
&=&
S(z,t)
+
\nabla_z(\kappa\nabla_z T_{\rm e})
-
G_{E_{2g}}(T_{\rm e}-T_{E_{2g}})
\nonumber\\
&&
- G_{\rm ph}(T_{\rm e}-T_{\rm ph}),
\label{eq:el}
\\
C_{E_{2g}}\frac{\partial T_{E_{2g}}}{\partial t}
&=&
G_{E_{2g}}(T_{\rm e}-T_{E_{2g}})
-
C_{E_{2g}}\frac{T_{E_{2g}}-T_{\rm ph}}{\tau_0},
\label{eq:e2g}
\\
C_{\rm ph}\frac{\partial T_{\rm ph}}{\partial t}
&=&
G_{\rm ph}(T_{\rm e}-T_{\rm ph})
+C_{E_{2g}}\frac{T_{E_{2g}}-T_{\rm ph}}{\tau_0}.
\label{eq:ph}
\end{eqnarray}
Here $C_{\rm e}$, $C_{E_{2g}}$, and
$C_{\rm ph}$
are the specific heat capacities for the electron,
hot-phonon, and cold-phonon states, respectively.
$G_{E_{2g}}$ ($G_{\rm ph}$)
is the electron-phonon relaxation rate between electronic states
and hot (cold) phonons modes,
calculated by
means of $\alpha^2F_{E_{2g}}$ ($\alpha^2F_{\rm ph}$).
%
%
Furthermore $\kappa$ is the thermal conductivity of electrons
and $\tau_0$ is a parameter ruling
the anharmonic phonon-phonon
scattering between the hot and cold phonon components
(for further details see Section\,S1 and Fig.\,S1 in Ref.\,\cite{bib:si}).
Modelling a typical pump-probe experiment
with the photon energy being $> 1$\,eV,
we assume the pump energy to be transferred
uniquely to the electronic DOFs
by the term $S(z,t)=I(t)e^{-z/\delta}/\delta$,
where $I(t)$ is the intensity of the absorbed fraction
of the laser pulse (with a Gaussian profile)
and $\delta$ is the penetration depth.
The anisotropic coupling of the e-ph interaction
is thus reflected in a different evolution of
the three characteristic temperatures.
Starting from an 
initial thermalized system at $T_0=300$\,K,
the energy pumped to the electronic DOFs is transferred faster to the
$E_{2g}$ phonons than to the other lattice vibrations,
leading to an effective temperature $T_{E_{2g}}$ significantly higher 
than that of the other modes, $T_{\rm ph}$. 
Final thermalization between all the lattice DOFs
occurs on time scales of several picoseconds, as a result
of the weak direct phonon-phonon scattering
and of the weak coupling between the electronic 
states and phonon modes other than the $E_{2g}$ ones.
In our calculations, the
parameters in Eqs.\,(\ref{eq:el})-(\ref{eq:ph})
(with the exception of $\kappa$, $\delta$ and $\tau_0$)
are evaluated numerically from the
first-principles calculations \cite{bib:si}.

\begin{figure}[!t]
\includegraphics[width=0.45\textwidth]{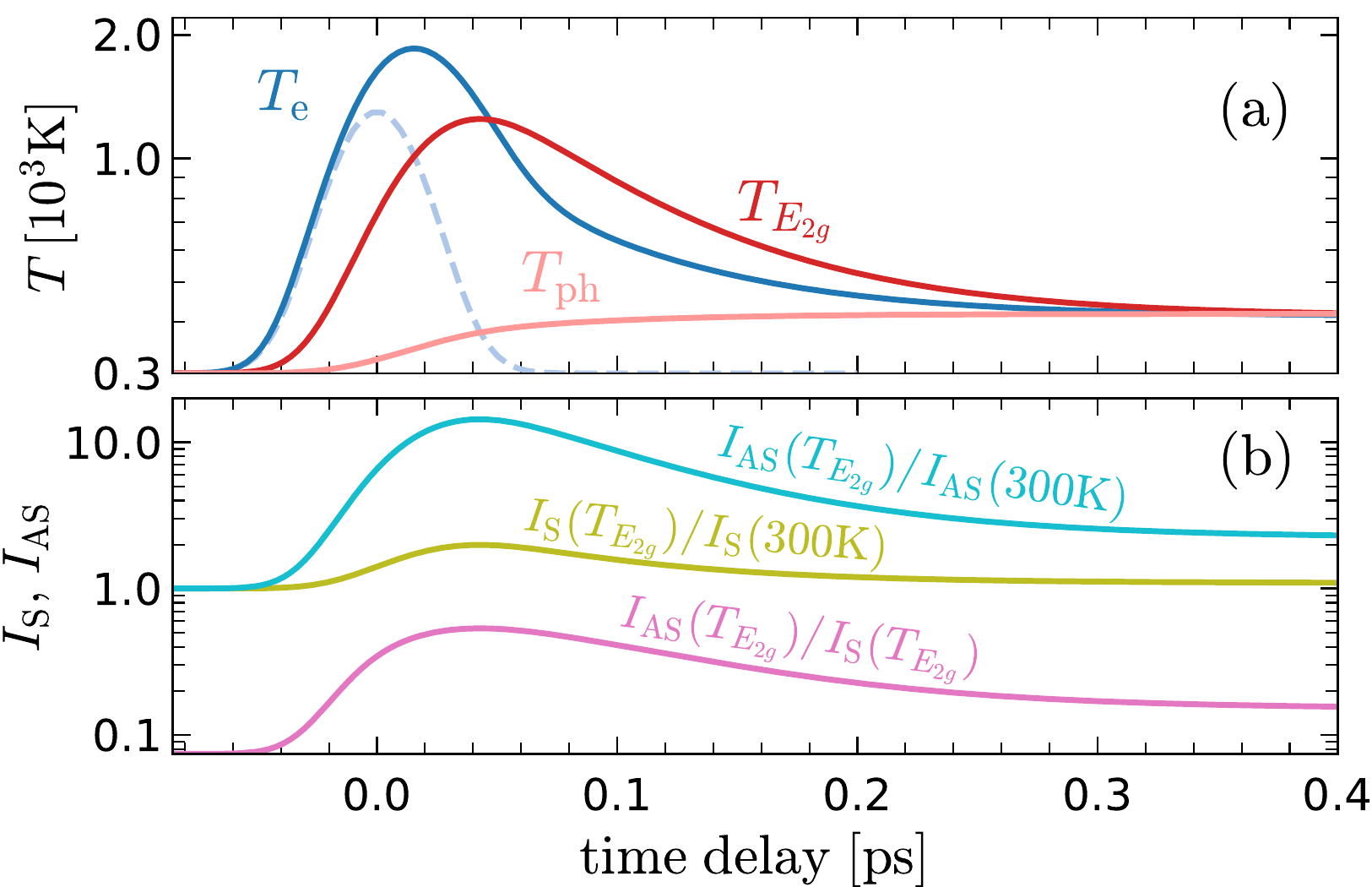}
\caption{\label{fig:fig2}(a) Time dependence of the electron and phonon effective 
temperatures $T_{\mathrm{e}}$, $T_{E_{2g}}$, $T_{\mathrm{ph}}$ in MgB$_2$ as obtained from the 
three-temperature model. The dashed line shows the pulse profile.
The absorbed fluence of the pump pulse is 12\,J/m$^2$, 
the pulse duration is 45\,fs (as in Ref.\,\cite{baldini}). 
(b) Ratios between the intensities of the Stokes ($I_{\rm S}$) 
and anti-Stokes ($I_{\rm AS}$) $E_{2g}$ Raman peaks. 
}
\end{figure}

Our calculations
predict a very fast increase of $T_{E_{2g}}$ [see Fig.\,\ref{fig:fig2}(a)],
reaching the maximum temperature
$T_{E_{2g}}^{\rm max}\approx 1200$\,K with a short delay
of $40$\,fs from the maximum
energy transfer to the electronic DOFs,
consistent
with a computed relaxation time $\tau_{E_{2g}}\approx 46$\,fs
(see Section S1 in Ref.\,\cite{bib:si}).
Subsequent thermalization between electrons, hot $E_{2g}$ phonons,
and the remaining lattice DOFs occurs
on a quite longer time scale, $\sim 1$\,ps \cite{bib:si}, where
all the DOFs thermalize to
an average temperature $\sim 400$\,K\,\footnote{
Note that the time evolution of $T_{\rm e}$ resembles
remarkably the two-$\tau$ behavior discussed in Ref.\,\cite{demsar03},
where, however, the idea of a fast electron-electron
thermalization was rejected.
The outstanding rise of $T_{E_{2g}}$ is not expected
to depend substantially on the effective electron-electron
thermalization (see Ref.\,\cite{bib:si}).
}.
Note that the strong enhancement of $T_{E_{2g}}$ with respect
to $T_{\rm ph}$ is not so much due to the difference
between $\lambda_{E_{2g}}$ and $\lambda_{\rm ph}$,
but
rather due to the smaller heat capacity $C_{E_{2g}} \ll C_{\rm ph}$,
reflecting the fact that very few $E_{2g}$ modes
in $\alpha^2F_{E_{2g}}$ are responsible
for a similar coupling as many cold lattice modes in 
$\alpha^2F_{\rm ph}$.

The preferential energy transfer to a single phonon mode
can be revealed via several experimental techniques.
One of the most direct ways is measuring the intensities
of the Stokes (S) and anti-Stokes (AS) $E_{2g}$ peaks in Raman spectroscopy, 
which are related to the Bose-Einstein occupation factor
$b(\omega;T)=[\exp(\omega/T)-1]^{-1}$ via 
the relations $I_{\rm S}(T_{E_{2g}})\propto
1+b(\omega_{E_{2g}};T_{E_{2g}})$ and 
$I_{\rm AS}(T_{E_{2g}})\propto
b(\omega_{E_{2g}};T_{E_{2g}})$, respectively.
Assuming to work at zero fluence and room temperature,
we predict in Fig.\,\ref{fig:fig2}(b) an increase of the intensity of the Stokes peak
up to a factor 2 [$I_{\rm S}(T_{E_{2g}})/I_{\rm S}(300\,\mbox{K})
\approx 2$], and of the anti-Stokes peak
as high as a factor 15
[$I_{\rm AS}(T_{E_{2g}})/I_{\rm AS}(300 \,\mbox{K})\approx 15$].
At the maximum temperature of the hot phonon,
the intensity of the anti-Stokes resonance
can be as high as 50\% of the intensity
of the Stokes peak.
The experimental investigation of Stokes and anti-Stokes peak intensities in time-resolved
Raman spectroscopy may  provide also
a {\em direct} way to probe the validity of the
hot-phonon scenario by simultaneous measurement of the Stokes/anti-Stokes
intensities of the Raman active out-of-plane $B_{1g}$ mode
with frequency $\omega_{B_{1g}}\approx 86$\,meV.
Since this mode is weakly
coupled to the electronic states, we expect it
to be governed by the cold-phonon temperature $T_{\rm ph}$,
with a drastically different behavior in the time evolution of the 
Stokes/anti-Stokes peak intensities than
the $E_{2g}$  mode (see Section S3 and Fig.\,S4 in Ref.\,\cite{bib:si}). 
These spectral signatures constitute a clear fingerprint of hot-phonon 
physics, suggesting that time-resolved Raman measurements may provide a 
tool to unambiguously unravel the thermalization mechanisms for systems 
out of equilibrium. 

As shown in Refs.\,\cite{bib:yan09,bib:ferrante18}, the peculiar characteristics
of hot-phonon dynamics can be traced
also through the $\omega$-resolved phonon spectral properties.
On the theoretical side, these properties can be properly investigated
in the Raman spectra of the $E_{2g}$ mode 
upon computation of the many-body phonon self-energy
$\Pi(\omega;\{T\})$
of the $E_{2g}$ mode at ${\bf q}\approx 0$\,\cite{bib:lazzeri06}.
Note that, in the real-time dynamics,
the phonon self-energy will depend on the {\em full}
set of electron and phonon temperatures
$\{T\}=(T_{\rm e}, T_{E_{2g}}, T_{\rm ph})$.
The full spectral properties can be thus evaluated
in terms of the phonon spectral function as \cite{bib:giustino17}:
\begin{align}
B(\omega;\{T\})
=
-
\frac{1}{\pi}
\mathrm{Im}
\left[
\frac{2\omega_{E_{2g}}}{\omega^2-\omega_{E_{2g}}^2
-2\omega_{E_{2g}}\overline{\Pi}(\omega;\{T\})}
\right],
\label{eq:phononspectral}
\end{align}
where $\omega_{E_{2g}} =  67$\,meV is the harmonic adiabatic
phonon frequency as obtained from DFPT
and $\overline{\Pi}(\omega;\{T\})$
is the phonon self-energy for the $E_{2g}$ modes,  where, to avoid double-counting,
the non-interacting adiabatic contribution at $T=0$\,K is subtracted
(for more details on the nonadiabatic phonon self-energy see Ref.\,\cite{bib:si}).
The inclusion of many-body effects 
on the crystal-lattice dynamics via Eq.\,(\ref{eq:phononspectral}) 
is reflected by renormalization of the phonon energy $\Omega_{E_{2g}}$
and by the finite phonon linewidth $\Gamma_{E_{2g}}$, which may be computed 
through solution of the following self-consistent equations: 
$\Omega_{E_{2g}}^2=\omega_{E_{2g}}^2
+2\omega_{E_{2g}}\overline{\Pi}(\Omega_{E_{2g}};\{T\})$,
and $\Gamma_{E_{2g}}=-2\mbox{Im}\overline{\Pi}(\Omega_{E_{2g}};\{T\})$.

\begin{figure}[t]
\includegraphics[width=0.48\textwidth]{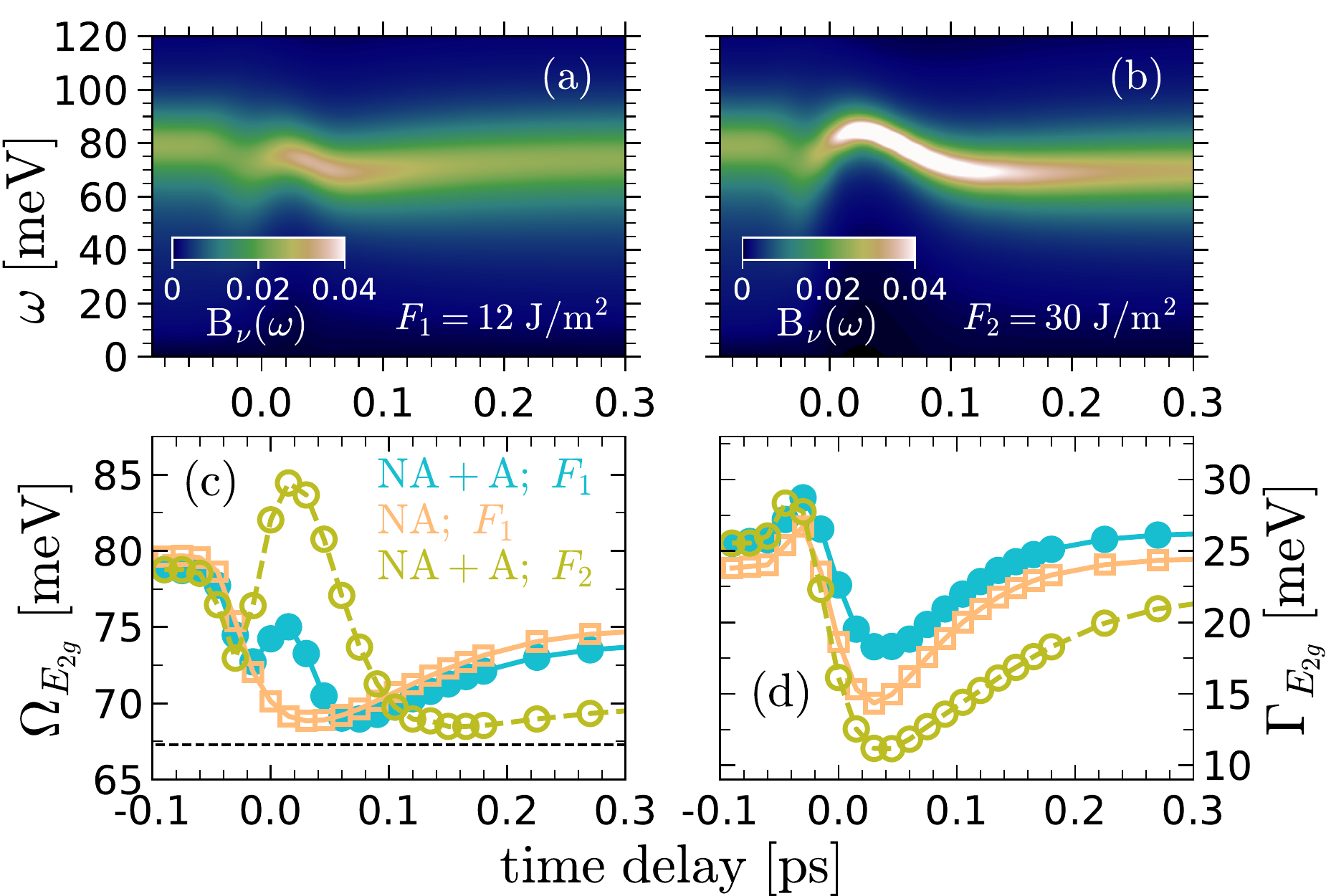}
\caption{(a)-(b) Intensity of the phonon spectral function
$B_{E_{2g}}(\omega;\{T\})$ for $F=12$\,J/m$^2$ (panel a)
and for $F=30$\,J/m$^2$ (panel b).
Time evolution of the (c) Raman peak positions
and (d) phonon linewidths using the full self-energy
for $F=12$\,J/m$^2$ (full circles) and
for $F=30$\,J/m$^2$ (open circles).
Also shown are the results obtained with only
the NA intraband term and for $F_2=30$\,J/m$^2$ 
(open squares).
The dashed horizontal line in panel (c) shows the
adiabatic energy of the $E_{2g}$ mode.
} 
\label{f-spectra}
\end{figure}

Using such theoretical tools,
we evaluate, within the three-temperature model,
the time-resolved dynamics
of the Raman peak position and of the phonon linewidth,
as well as of the full phonon spectral function of the $E_{2g}$ mode in MgB$_2$
as a function of the pump-probe time delay.
A similar approach (however, without time dependence) was used
in Ref.\,\cite{bib:ferrante18}
for graphene, where
the effects of the electronic damping
due to the electron-electron interaction was explicitly included
in the evaluation of the phonon self-energy.
This description is however insufficient in the case of MgB$_2$
where the electronic damping is crucially governed
by the e-ph coupling itself\,\cite{bib:cappelluti06,novko18}.
In order to provide a reliable description we evaluate thus
the $E_{2g}$ phonon self-energy 
in a nonadiabatic framework\,\cite{novko18}
explicitly 
retaining the e-ph renormalization effects in the Green's functions of
the relevant intraband contribution
(see Section S4 in Ref.\,\cite{bib:si}).
The $E_{2g}$ phonon spectral function is  shown
in Fig.\,\ref{f-spectra}(a)-(b) as function of the time delay,
for two different fluences.
The corresponding phonon energies
$\Omega_{E_{2g}}$ and linewidths $\Gamma_{E_{2g}}$
are summarized
in panels (c) and (d).
%
The combined effect of the time evolution of $T_{\rm e}$
and $T_{E_{2g}}$, $T_{\mathrm{ph}}$ results in
a non-trivial time-dependence
of the spectral properties.
Our calculations reveal a counter-intuitive
{\em reduction} of the phonon linewidth $\Gamma_{E_{2g}}$ 
right after photo-excitation, followed by a subsequent increase
during the overall thermalization with the cold phonon DOFs.
The time dependence of the phonon frequency shows an even more
complex behavior, with an initial redshift, followed by a partial
blueshift, and by a furthermore redshift.

In order to rationalize these puzzling results, we analyze in detail
the temperature dependence of the phonon spectral properties,
decomposing the phonon self-energy
in its basic components:  interband/intraband terms,
and in adiabatic (A) and nonadiabatic (NA) processes.
For details see Ref.\,\cite{bib:si},
whereas here we summarize the main results.
A crucial role is played by the NA intraband term, which 
is solely responsible for the phonon damping.
Following a robust scheme usually employed for the optical conductivity
(see Section S4 in Ref.\,\cite{bib:si}),
we can model the 
effects of the e-ph coupling on the intraband
processes in terms of the renormalization function
$\lambda(\omega;\{T\})$ and the
e-ph particle-hole scattering rate
$\gamma(\omega;\{T\})$:
\begin{align}
\overline{\Pi}^{\rm intra, NA}(\omega;\{T\})
=
\frac{\omega \langle|g_{E_{2g}}|^2\rangle_{T_e}}
{\omega[1+\lambda(\omega;\{T\})]+i\gamma(\omega;\{T\})},
\end{align}
where
$\langle|g_{E_{2g}}|^2\rangle_{T_e}=
-
\sum_{n\mathbf{k}\sigma}
\left|g_{E_{2g}}^{nn}(\mathbf{k})\right|^2
\partial f(\varepsilon_{n\mathbf{k}};T_{\rm e})/
\partial \varepsilon_{n\mathbf{k}}$ \cite{bib:si}.
Phonon optical probes at equilibrium
are commonly at room (or lower) temperature
in the regime $\gamma(\omega;T)
\ll \omega[1+\lambda(\omega;T)]$, where the
phonon damping $\Gamma_{E_{2g}} \propto \gamma(\Omega_{E_{2g}};T)$.
Our calculations predict on the other hand
 $\gamma(\Omega_{E_{2g}};T_{300\mathrm{K}})\approx
75$\,meV, which is close to
$\Omega_{E_{2g}}[1+\lambda(\Omega_{E_{2g}};T_{300\mathrm{K}})]\approx 85$\,meV,
resulting in $\Gamma_{E_{2g}} \approx 26$\,meV, in good agreement
with the experiments\,\cite{hlinka,goncharov,bib:martinho03} and with the previous calculations\,\cite{bib:cappelluti06,novko18}.
The further pump-induced increase of
$\gamma(\Omega_{E_{2g}};\{T\})\gg
\Omega_{E_{2g}}[1+\lambda(\Omega_{E_{2g}};\{T\})]$
drives the system into an opposite regime where
$\Gamma_{E_{2g}} \propto 1/\gamma(\Omega_{E_{2g}};T)$.
In this regime the pump-induced increase of
$\gamma(\Omega_{E_{2g}};\{T\})$ results thus
in a {\em reduction} of $\Gamma_{E_{2g}}$,
as observed in Fig.\,\ref{f-spectra}(d).
A similar change of regime is responsible for
the crossover from an Elliott-Yafet to the Dyakonov-Perel
spin-relaxation, or for the NMR motional
narrowing \cite{boross13,szolnoki17}.
We also note here that the same effects and the change of regime
are partially responsible for the overall time-dependence
of the phonon frequency [see Fig.\,\ref{f-spectra}(c)],
where the full result (full blue circles) is compared with the one retaining only
the nonadiabatic intraband self-energy (open orange squares).
The redshift predicted for the the latter case is a direct effect
of the same change of regime responsible for the reduction
of the phonon damping.
However, in the real part of the self-energy, adiabatic processes
(both intra- and inter-band)
play also a relevant role \cite{bib:si}, giving rise
to an additional blueshift (ruled uniquely by $T_{\rm e}$)
that partially competes with the redshift induced by nonadiabatic
intraband processes.
Note that actual magnitude of this anomaly
depends on the pump fluence [compare full and open circles
in Fig.\,\ref{f-spectra}(c)].
This dependence can be also used to trace down such adiabatic processes.
For a realistic possibility of detecting these spectral features
in time-resolved Raman spectroscopy one needs to face
the limitations of the time-energy uncertainty \cite{versteeg18}.
For a time resolution of $\sim 50$\,fs, comparable with the pulse width,
one gets a energy resolution of 
$\sim 36$\,meV.
While this limitation would prevent the detection of fine structures,
the coherent shift of the peak center and the time-dependence of the phonon linewidth
should be clearly observable\,\cite{fausti}
(see also Fig.\,S5 shown in Ref.\,\cite{bib:si}).
Furthermore, the development of alternative techniques based on
quantum/statistical correlations\,\cite{randi17,tollerud19}
has shown to provide a promising way to overcome
the limitations of the time-energy uncertainty.
Therefore, the insights given here along with our \emph{ab initio}
method might be of general importance, especially considering that
the theoretical framework for deciphering ultrafast phonon dynamics
is at the moment not present in the literature.

In conclusion, in this Letter we have presented
a quantitative and compelling evidence that
a hot-phonon scenario dominates 
the ultrafast carrier dynamics of MgB$_2$ in
time-resolved pump-probe experiments.
We further predict the emergence of specific spectral signatures in 
time-resolved Raman spectroscopy, which may 
guide the direct experimental verification of a hot-phonon regime in MgB$_2$.
The present analysis is of interest for understanding
and controlling the coupling mechanisms in this material,
with further relevance for technology.
Possible future applications can range
from optical probes for sensoring the internal temperature
to controlling the heat transfer
between electronic and lattice DOFs
in order to optimizing dissipation
processes and interfaces between superconducting
and normal metals.

\begin{acknowledgments}
We thank F. Carbone, E. Baldini, L. Benfatto, D. Fausti, A. Perucchi 
and P. Postorino for enlightening discussions.
D.N. gratefully acknowledges financial support from the 
European Regional Development Fund for the ``Center of 
Excellence for Advanced Materials and Sensing Devices'' 
(Grant No. KK.01.1.1.01.0001). Financial support by Donostia 
International Physics Center (DIPC) during various stages of 
this work is also highly acknowledged. Computational resources 
were provided by the DIPC computing center.
\end{acknowledgments}

\bibliography{ufmgb}

\end{document}